\documentclass{aa}
\usepackage{array}
\usepackage{tabularx}
\usepackage{txfonts}
\usepackage{natbib}
\usepackage{graphicx}
\usepackage{longtable}

\newcommand{\hd}{\object{HD 141569}}

\makeatletter
\def\cbrace#1{\@cbrace#1\@nil}
\def\@cbrace#1-#2\@nil{%
  \omit
  \@multicnt#1%
  \advance\@multispan\m@ne
  \ifnum\@multicnt=\@ne\@firstofone{&\omit}\fi
  \@multicnt#2%
  \advance\@multicnt-#1%
  \advance\@multispan\@ne
  \downbracefill
  \cr
  \noalign{\vskip-2\arrayrulewidth}}
\makeatother
\begin{document}
   \title{Investigating the flyby scenario for the HD 141569 system}

   \author{R. Reche
          \and
          H. Beust
          \and
          J.-C. Augereau
          }

   \offprints{R. Reche \email{Remy.Reche@obs.ujf-grenoble.fr}}

   \institute{Laboratoire d'Astrophysique de Grenoble, CNRS,
     Université Joseph Fourier,
UMR 5571, Grenoble, France}

\date{Received <date>; accepted <date>}

\abstract
    {\hd, a triple star system, has been  intensively observed and
studied for its massive debris disk. Until recently, it was rather
   regarded as a gravitationally bound triple system but
   recent measurements of the HD 141569A radial velocity seem to
   invalidate this hypothesis. The flyby scenario has therefore to be
   further investigated to test
   its compatibility with the observed disk structures.}
    {We present a study of the flyby scenario for the HD141569 system,
   by considering $3$ variants: a sole flyby, a flyby associated with
   one planet embedded in the disk and a flyby with two planets in the
   disk. We discuss the merits of each one to reproduce the scattered
   light observations of the disk.}
    {We first use analytical calculations to reduce the parameter space
   of the $2$ stellar companion's orbit and then perform N-body numerical
   simulations of the flyby encounter, using symplectic integration,
   taking into account the gravitational influence of  
   the stars and the planets on massless test particles.}
    {The binary orbit is found to be almost fixed by the
   observational constraint on a edge-on plane with respect to the
   observers. If the binary has had an influence on the disk
   structure, it should have a  passing time at the periapsis between $5000$
   and $8000$ years ago and a distance at periapsis between $600$ and
   $900$ AU. It also appears that the best scenario for reproducing
   the disk morphology is a
   flyby with $1$ planet embedded in the disk. For a $2$ $M_J$
   planet, its orbital eccentricity must be around $0.2$ while for a $8$
   $M_J$ planet, it must be below $0.1$. In the two cases, its
   apoapsis is about $130$ AU.}
    {Although th global disk shape is reasonably well reproduced, some
   observed features cannot be explain by the present model and the
   likehood of the flyby event remains an issue for the scenario
   explored in this paper. Dynamically speaking, HD 141569 is still a
   puzzling system.}


   \maketitle
%

\section{Introduction}
\hd, a triple star system, has been intensively observed and
studied for its circumstellar dusty disk imaged in scattered light in
the visible and at near-infrared wavelengths. This system is located at
$99\pm10$ pc and its age is estimated to $5\pm3$ Myrs
\citep{2000ApJ...544..937W,2004A&A...419..301M}.  The disk is
associated to the central star (B9.5 Ve) while the two other
companions (M2 and M4) form a binary.

The resolved images of the disk, showing a complicated
morphology, have generated a lot of discussion in the literature
about the systems dynamics (see Table \ref{previousWorks} for a
summary). First of all, this disk is  in transition
to  a debris disk:
although it is a dusty optically thin disk with a fractional disk
luminosity of $L_{disk}/L_* \simeq 8.4 \times 10^{-3}$ \citep{1996MNRAS.279..915S},
the gas mass is not negligible. It could in fact represent most of the
mass according to \citet{2006A&A...453..163J} ($M_{gas}=80 \, M_{\oplus}$,
$M_{dust}=2.2 \, M_{\oplus}$ for grains with radius sizes between $1$ $\mu
$m and $1$ cm), although the spatial distribution of the
gas and of the dust may differ. The
influence of the gas on the dust dynamics has been therefore
taken into account by \citet{2005ApJ...627..986A}, but
other authors consider only
classical N-Body simulations without gas
\citep{2004AA...414.1153A,2005DPS....37.2810W}.
\citet{2005AJ....129.2481Q}, on the other hand, only considered
the gas in their simulations.

The dynamical status of the
external binary is also questionable: in order to reproduce the
external structures of the disk, previous studies have either considered
the case of gravitational bound companions
\citep{2004AA...414.1153A,2005AJ....129.2481Q} or the case of a
flyby \citep{2005ApJ...627..986A}. The internal disk structures might be explained by
the gravitational perturbation of an unseen planet at large distance,
around $250$ AU
\citep{2005DPS....37.2810W} but also by  alternative
mechanisms, such as dust migration in a gaseous disk
\citep{2001ApJ...557..990T}. The latter scenario is successful in
producing annular structures, but cannot account for the observed
non-axisymmetric of these features. The
combined effect of a 
planet and  external perturbers has only been considered in one
study, without successfully reproducing all the structures
\citep{2005ApJ...627..986A}. 

\begin{figure*}
\centering
\makebox[\textwidth]{
\includegraphics[angle=-90,width=0.33\textwidth]{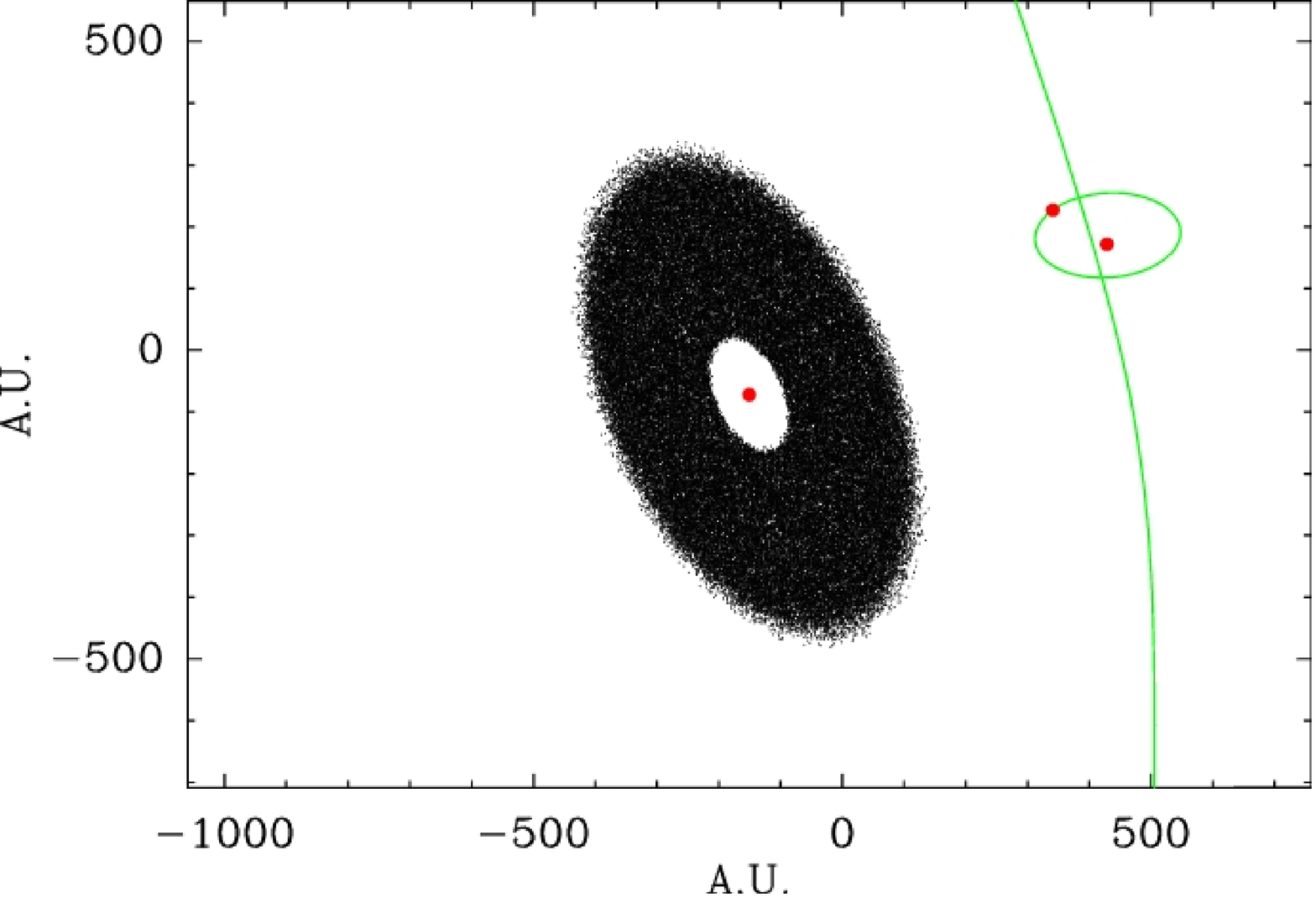}\hfil
\includegraphics[angle=-90,width=0.33\textwidth]{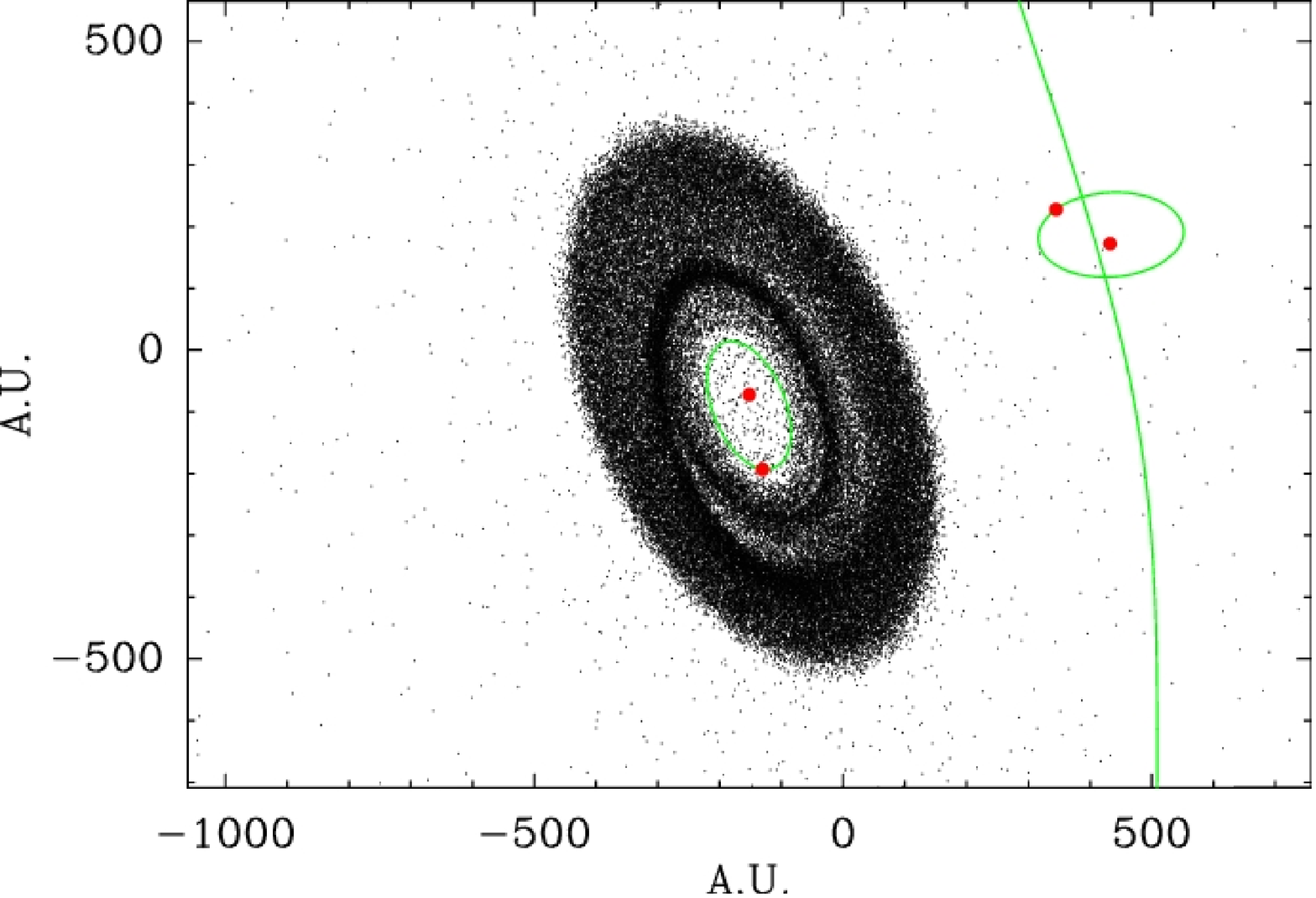} \hfil
\includegraphics[angle=-90,width=0.33\textwidth]{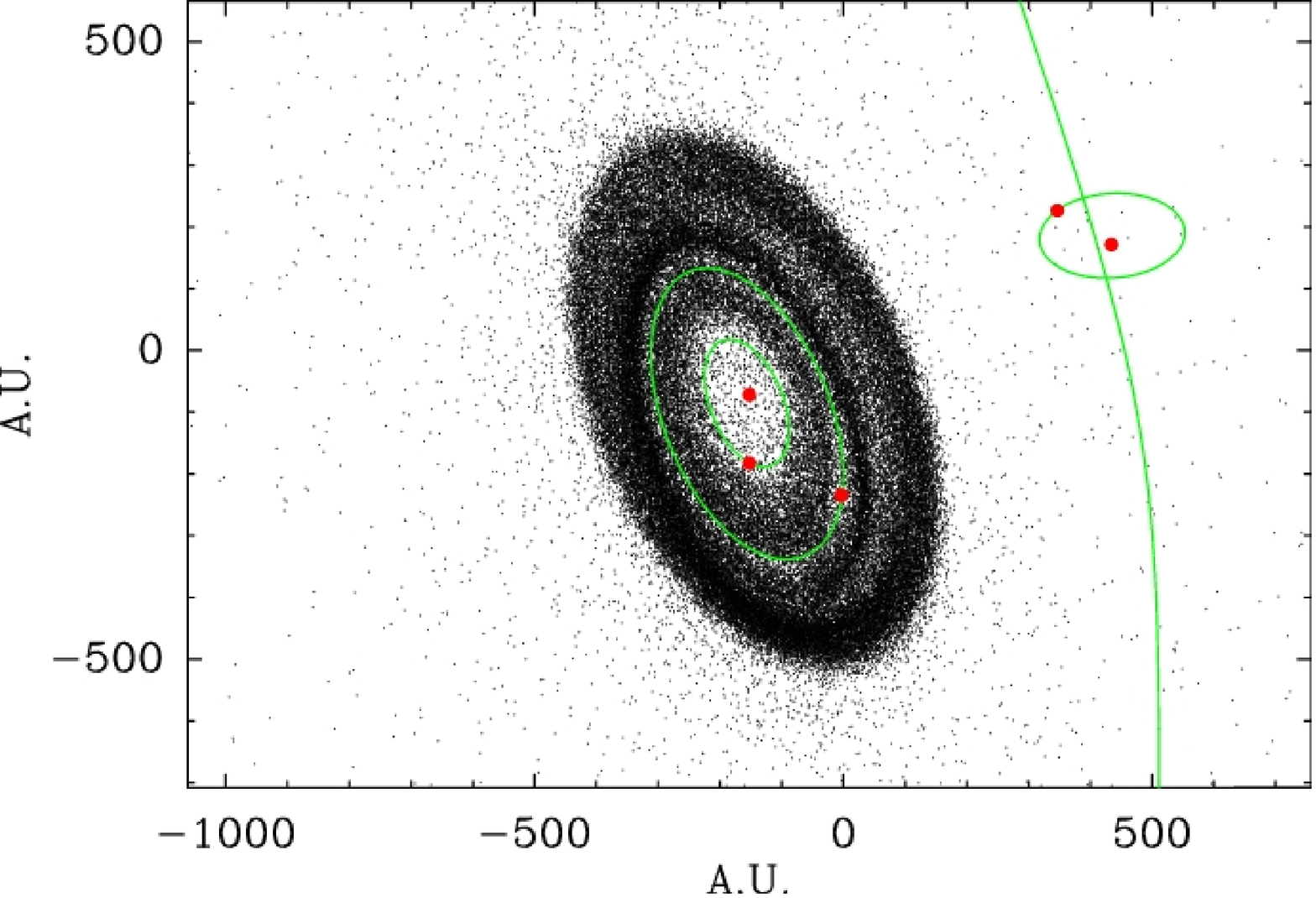}} \\
\caption{\label{sketch} Illustrations of the $3$ scenarios studied in
    this paper: a sole flyby (left panel), a flyby + $1$ planet
    (middle panel) and a flyby + $2$ planets (right panel). The stars
    and planet locations, are represented by large red points, and the
    various orbits  by a thin green line. \thanks{See the electronic edition of the
    Journal for a color version of this figure.}}
\end{figure*}

New millimeter observations of the gas disk (\citet{2005MNRAS.359..663D} and Augereau
et al., in prep) , which give better constraints on the
radial velocity of the primary star, show that the
differential velocity between the binary and the central star  ($5.8
\pm 0.3$ km s$^{-1}$) is
larger than the system escape velocity ($2.6$ km s$^{-1}$) so that the flyby scenario seems
therefore to be the most plausible. Therefore, we propose in this paper to extensively study the flyby
scenario, as it is the least studied up to now. In
Sect. \ref{observations} we summarise the available observations of
the disk and the stars astrometry constraints. In order to best reproduce the
observations, we discuss $3$ different
scenarios (Fig. \ref{sketch}): a sole flyby
(Sect. \ref{constraints}), a flyby with $1$ planet embedded in the disk
(Sect. \ref{planet})  and a flyby with $2$ planets embedded in the disk
(Sect. \ref{2planets}). In Sect. \ref{discussion}, we compare our
approach to previous studies and discuss in particular the likelihood of
this scenario. We finally summarise and conclude in Sect. \ref{conclusion}.

\begin{table*}
\centering
\caption{\label{previousWorks}Summary of recent papers on the
  dynamical modelling of the \hd  system.}
\begin{tabular*}{\textwidth}{@{\extracolsep{\fill}} c c c c c}
Authors & Modelled structures & External companions & Planet & Notes \\
\hline\hline
 \citet{2001ApJ...557..990T} & gap \& two rings & No & No & Dust
 migration due to gas friction \\
\citet{2004AA...414.1153A} & outer ring & bound & No & N-Body gravitational code \\
\citet{2005AJ....129.2481Q} & outer ring & bound & No & 2D
  hydrodynamics code \\
\citet{2005ApJ...627..986A} & spiral arms \& inner depletion & flyby & Yes &
N-body code, SPH code, collisions \\
\citet{2005DPS....37.2810W} & gap \& outer ring & No & Yes & N-body
gravitation code \\
\hline
\end{tabular*}
\end{table*}

\section{Available Observations}
\label{observations}
\subsection{Overview of the disk shape}
The motivation for our study is to reproduce the dust disk structures
that have been
observed in scattered light. According to the visible, near-infrared
and mid-infrared observations from 
\citet{1999A&A...350L..51A}, \citet{1999ApJ...525L..53W},
  \citet{2001A&A...372L..61M}, \citet{2002ApJ...573..425M} and  \citet{2003AJ....126..385C}
the following facts appear:
\begin{itemize}
\item The main structure is the two rings shape of the disk. The two
  annuli peak at $\sim 200$ AU  and $\sim 325$
  AU from the star. The two bright rings show out of phase brightness asymmetries of
  up to factors of $2.5$-$3$ for the outer ring in  the visible. The outer ring
  moreover shows a tightly-wound spiral structure. The two annuli are
  separated by a darker ring or ``gap'', which is radially wide
  compared to the two annuli.
\item an extended diffuse emission associated with a faint spiral arm is present in the North-East of
  the disk and is detected up to more than $600$ AU. According to the
  millimeter maps and using the hypothesis of forward
  scattering, this spiral would be a trailing structure with respect to the
  disk rotation. An other spiral arm is possibly observed, pointing
  toward the companions.
\item the disk brightness sharply decreases between $200$ and $150$ AU
  from the central star
  until the background level is reached. This behaviour is suggestive of a strong
  depletion of dust  inside $150$ AU.
\end{itemize}
\subsection{Radial velocities \& astrometry}
\citet{2000ApJ...544..937W} have measured the radial velocities of the
\hd B and C stars, respectively  $-1.5 \pm 0.6$ km s$^{-1}$ and  $-2.4 \pm
0.7$ km s$^{-1}$ by cross-correlating their spectra to that
of standard stars. This method was not suitable for HD 141569A due to its lack of
lines for comparison with the radial velocity
standard. \citet{2000ApJ...544..937W} also summarise the astrometry of
the $3$ stars from $1938$ to $1999$.

Augereau et al. (in prep) and \citet{2005MNRAS.359..663D} have measured the
double-peaked CO J=2-1 spectral profile of the disk around HD
141569A. It is thus possible to deduce  the radial velocity of this
star from the CO line and to obtain a far smaller uncertainty on this
measure. To a first approximation, assuming a symmetric emission in the
blue-shifted and red-shifted parts of the disk, the radial
heliocentric velocity of HD 141569A is found to be $-7.6 \pm 0.3$ km s$^{-1}$ (mean velocity
in the line). Supposing that the star masses are $2.5$, $0.5$
and $0.25$ $M_\odot$ for A, B and C respectively
\citep{2000ApJ...544..937W} and that the distance between HD141569A and the
center of mass of the B-C binary is at least $800$ AU (projected
distance, \citet{2000ApJ...544..937W}), the maximum escape velocity is $2.6$ km
s$^{-1}$. As the measured velocity is far above this limit, the two M
star companions are not bound to HD 141569A. In the context of
dynamical perturbation of the disk by the companions, the flyby scenario seems
therefore to be the most plausible. 

\section{The flyby scenario}
\label{constraints}
\subsection{Kinematic constraints}
\label{kinematics}
We define a cartesian coordinate system
$(X,Y,Z)$ where $X$ points toward the north, $Y$ toward the east and
$Z$ toward the Earth and where the origin is HD141569A. In this
referential frame, the position of the center of mass of the B-C
binary will be described by the coordinates $(X_b,Y_b,Z_b)$ and its
velocity by $V_{b_x},V_{b_y},V_{b_z}$. We already
know the present $X_b$, $Y_b$ (right ascension and declination)  and $V_{b_z}$ (radial velocity). But
the hyperbolic orbit of the binary around the primary
is  also  defined by the $6$ orbital
elements $(q,e,i,\omega,\Omega,u)$ where $q$ is the pericenter
distance, $e$ the eccentricity, $i$ the inclination with respect to
the sky plane ($OXY$), $\omega$ the
periapsis argument, $\Omega$ the longitude of the ascending node and
$u$ the eccentric anomaly at the observing time. The transformation from cartesian
coordinates to orbitals elements, or the inverse transformation, is
defined by a system of $6$ non-linear equations from classical
keplerian formalism. It is thus possible to look for orbital element sets verifying
the $3$ observables $(X_b,Y_b,Vz_b)$. But with $6$ unknown and only
$3$ constraints the equation system is underconstrained. One has
therefore to consider $3$ orbital
elements as free parameters in order to solve the system for the $3$
other orbital elements. This system is defined by the following
equations, for an hyperbolic orbit:

\begin{eqnarray}
\label{eq1}
X_b\cos{\Omega}-Y_b\sin{\Omega} & = & a\cos{i}\left(
 \sqrt{e^2-1}\cos{\omega}\sinh{u}\right. \nonumber \\
 && +  \left. \left( e-\cosh{u} \right) \sin{\omega} \right)\\
X_b\sin{\Omega}+Y_b\cos{\Omega} & = & a\left(
 -\sqrt{e^2-1}\sin{\omega}\sinh{u}\right. \nonumber \\
 && +  \left. \left( e-\cosh{u} \right) \cos{\omega} \right)\\
V_{b_z}\sqrt{a} \left( e\cosh{u} - 1 \right)&=&\sqrt{GM}\sin{i} \left(
 \sinh{u}\sin{\omega} \right. \nonumber \\
&&\left. -\sqrt{e^2-1} \cosh{u}\cos{\omega}  \right)
\end{eqnarray}

where $G$ is the gravitational constant and $M$ the sum of the $3$ star
masses. For this study, we choose to consider $(i,\Omega,u)$ as
free parameters and $(q,e,\omega)$ as unknowns. Once the $3$
parameters  $(i,\Omega,u)$ are
fixed to an arbitrary set of values, the non-linear system generally admits
none or one solution for the triplet $(q,e,\omega)$. For our problem,
many triplets of the parameters $(i,\Omega,u)$ give a
solution: for each $(i,u)$pair , at least one value of $\Omega$ gives
a valid solution for $(q,e,\omega)$. The results are thus not
significant because all the scenarios can be compatible with the observations:
orbital plane perpendicular to the line of sight, or instead very
inclined orbits, more or less  eccentric hyperbolic
orbits ... 

However, an other
constraint exists, not yet used in our method: the observed proper
motion of the stars
between $1938$ and $1998$ \citep{2000ApJ...544..937W}, and the associate
velocity $V_{sky} = 1.1 \pm 0.7$ km s$^{-1}$. The
uncertainty on this observable is quite large and we consider it as an
upper limit: all the solutions
that predict a proper motion less than twice the observed proper
motion are valid: 

\begin{equation}
\label{vsky}
\centering
\sqrt{V_{b_x}^2+V_{b_y}^2}<2V_{sky}
\end{equation}

\begin{figure}
\resizebox{\hsize}{!}{\includegraphics[angle=-90]{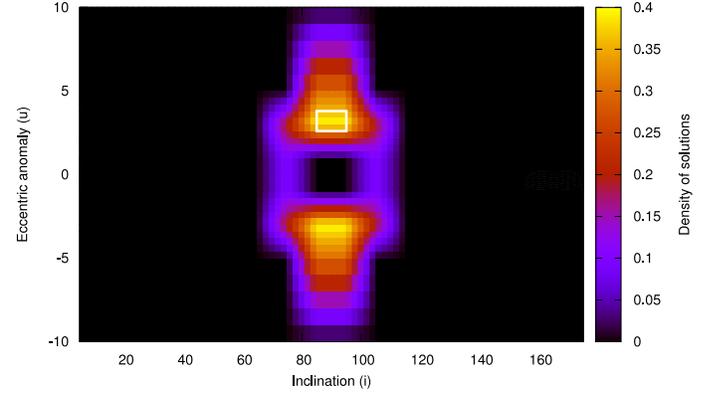}} 
\caption{\label{Solutions} Density of valid orbital configurations in
  the parameter space $(i,u,\Omega)$. The two axis are the eccentric
  anomaly $u$ and the inclination $i$,
  while the colors correspond to the density of the valid solutions for
  the longitude of the ascending node $\Omega$ in the range
  $[0,2\pi]$. The small white box shows the final valid parameter space
   when the disk geometry constraints are also taken in account
  (Sect. \ref{geometry}).\thanks{See the  electronic edition of the
    Journal for a color version of this figure.}}
\end{figure}

This is enough to reduce significantly the number of solutions, as
showed in Fig. \ref{Solutions}. On
this plot, the two axes are the eccentric anomaly $u$ and the
inclination $i$,
while the colors correspond to the density of valid solutions
(i.e. fulfilling the relation 1 to 4) for
the longitude of the ascending node $\Omega$ in the range $[0,2\pi]$. The
parameter space region where the valid solutions exist is indeed
very much reduced by using the proper motion constraint as we find,
with an
inclination $i$ between $75^\circ$ and $105^\circ$ and an eccentric
anomaly $u$ between $-5$ and $5$ radians. The fourth constraint has
therefore eliminated all the solutions with lower inclination and
larger proper motion.

The disk inclination is $55^\circ \pm 1.2^\circ$ from pole-on
\citep{2001A&A...372L..61M}, or rather $180^\circ - 55^\circ$ in order
to have the disk rotation \citep{2006ApJ...652..758G} consistent with
our convention (i.e. $i<90^\circ$ is an anticlockwise orbit on the
sky). The binary orbits with an inclination below
(resp. above) $90^\circ$ are thus retrograde (resp. prograde) with
respect of the disk rotation. The  relative inclination $i_{rel}$ between the
orbital plane of the binary and that of the disk can be calculated by the
following equation:

\begin{equation}
\label{irel}
\cos{i_{rel}}=\cos{i}\cos{i_{disk}}+\sin{i}\sin{i_{disk}}
\cos{\left( \Omega-\Omega_{disk}\right) }
\end{equation}

with $\Omega_{disk}$, the position angle of the disk ($357^\circ \pm
2^\circ$) and $i$, $\Omega$ orbital elements of the binary orbit.

\subsection{Disk geometry constraints}
\label{geometry}
At this point, only the kinematic constraints have been used to
calculate  the valid orbital configurations for the binary. However,
the encounter geometry is also constrained by the observed structures
in the disk if one assumes that  the current disk shape is a consequence of the
gravitational perturbations of the disk by the binary. The comparison
study of numerical  simulations of the flyby
scenario to the observations therefore reduces even more the range of
valid orbital configurations as detailed below.

\subsubsection{Numerical model}
\label{model}
We consider in our simulations a hierarchical system consisting of a
central star associated with a debris disk and an external binary. The
trajectory of this binary can be a bound orbit around the
primary star or a flyby (hyperbolic orbit). We address the case of
a disk consisting of large  particles (planetesimals), which are insensitive to
pressure forces (radiation, stellar wind or gas
pressure). Importantly, we also do not take into account the
gravitational  interactions
between  planetesimals as they are negligible, nor mutual
collisions. Dynamically speaking, the planetesimals are thus
considered as test particles. In Sect. \ref{planet} and \ref{2planets}, one or several
planets will be added to the simulations, orbiting the
central star in the orbital plane of the disk. The central star is
assumed to have a mass of $2.5$ M$_\odot$ and the low mass companions companions $0.5$ M$_\odot$ and
$0.25$ M$_\odot$ \citep{2000ApJ...544..937W}. The initial
disk consists of $100\,000$ test particles with a surface density
distribution  proportional to
$r^{-1}$. 

To perform our simulations we use the symplectic package
HJS \citep{2003A&A...400.1129B}, a SWIFT variant \citep{1991AJ....102.1528W,1994Icar..108...18L}
for hierarchical systems. It allows us to integrate accurately and
fastly the
motion of the disk particles, the relative motion of the two
companions and their orbit around the central star,
although they are on different timescales. Numerous scenarios can be
therefore tested and studied in a reasonable CPU time.

\subsubsection{Results}
\begin{figure*}
\centering
\makebox[\textwidth]{
\includegraphics[angle=-90,width=0.48\textwidth]{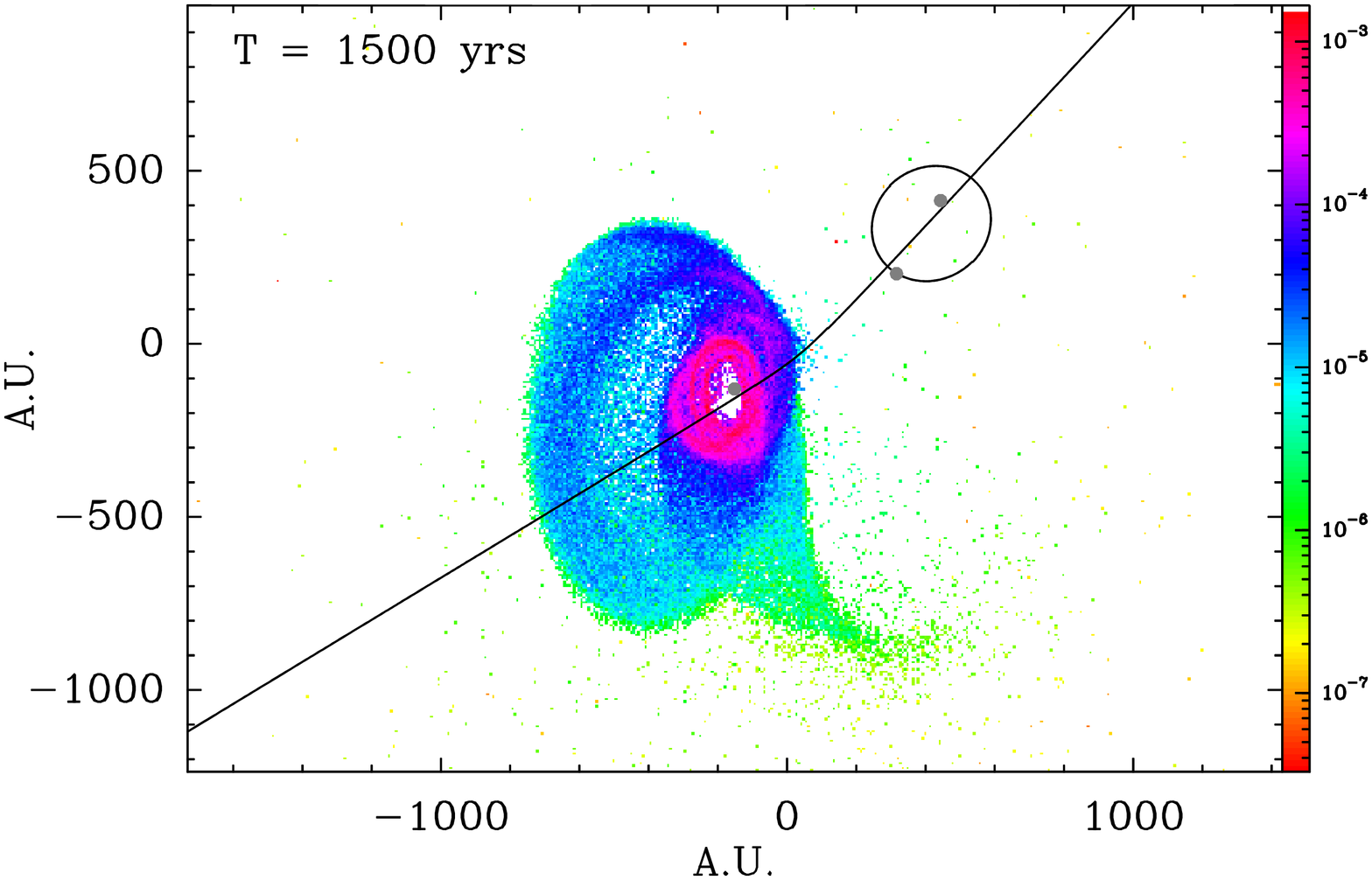}\hfil
\includegraphics[angle=-90,width=0.48\textwidth]{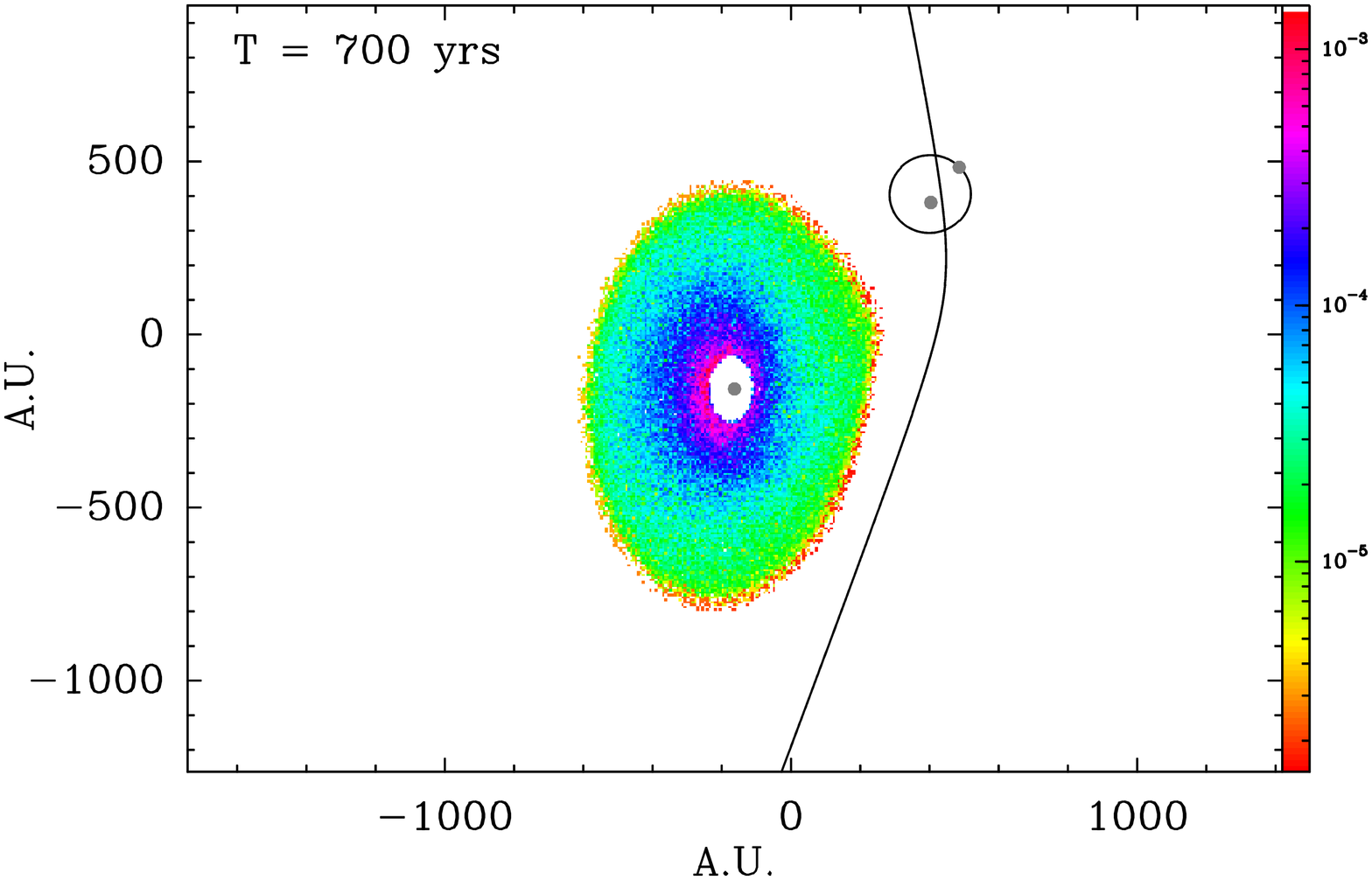}}\\
\makebox[\textwidth]{
\includegraphics[angle=-90,width=0.48\textwidth]{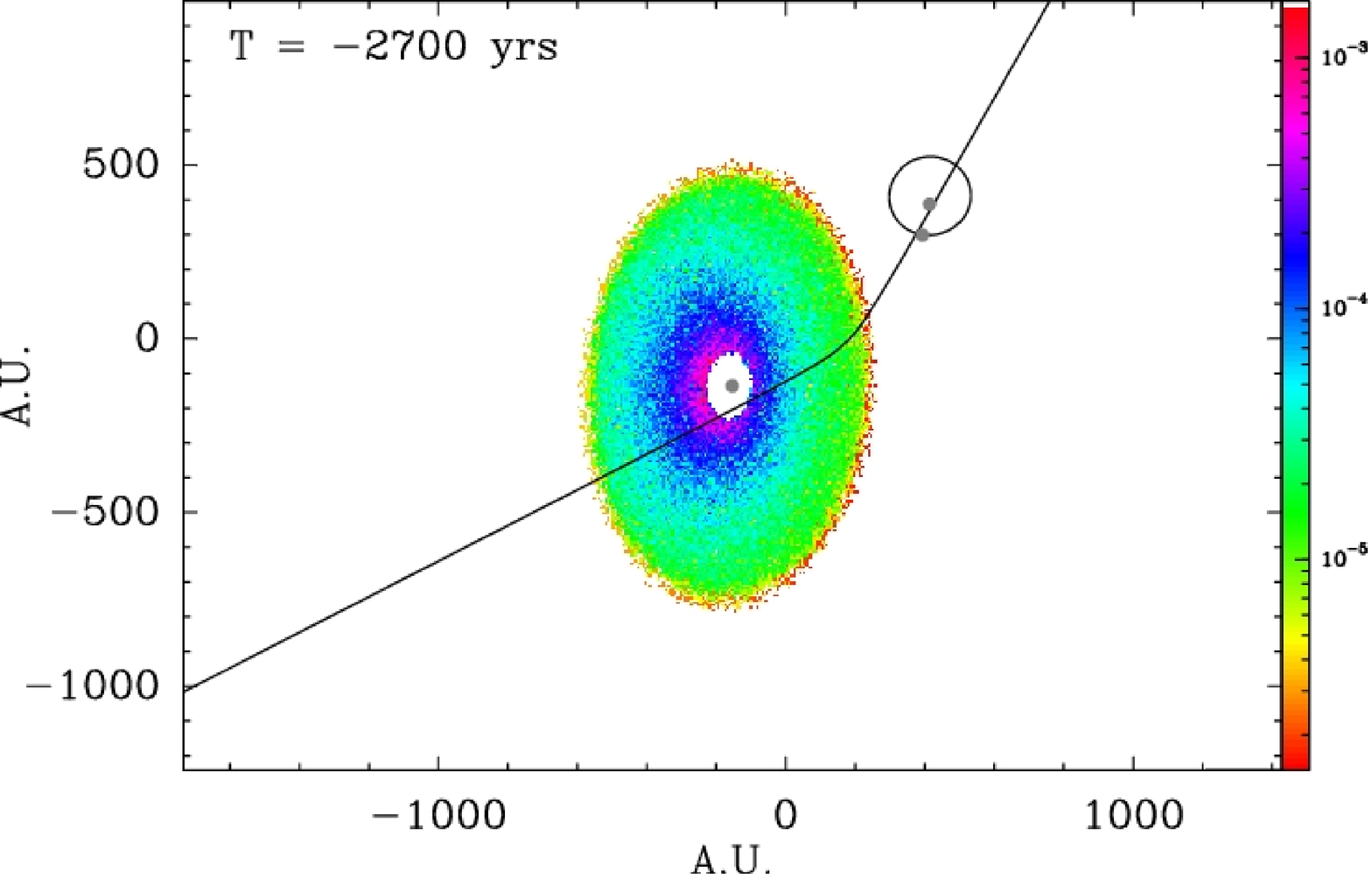}\hfil
\includegraphics[angle=-90,width=0.48\textwidth]{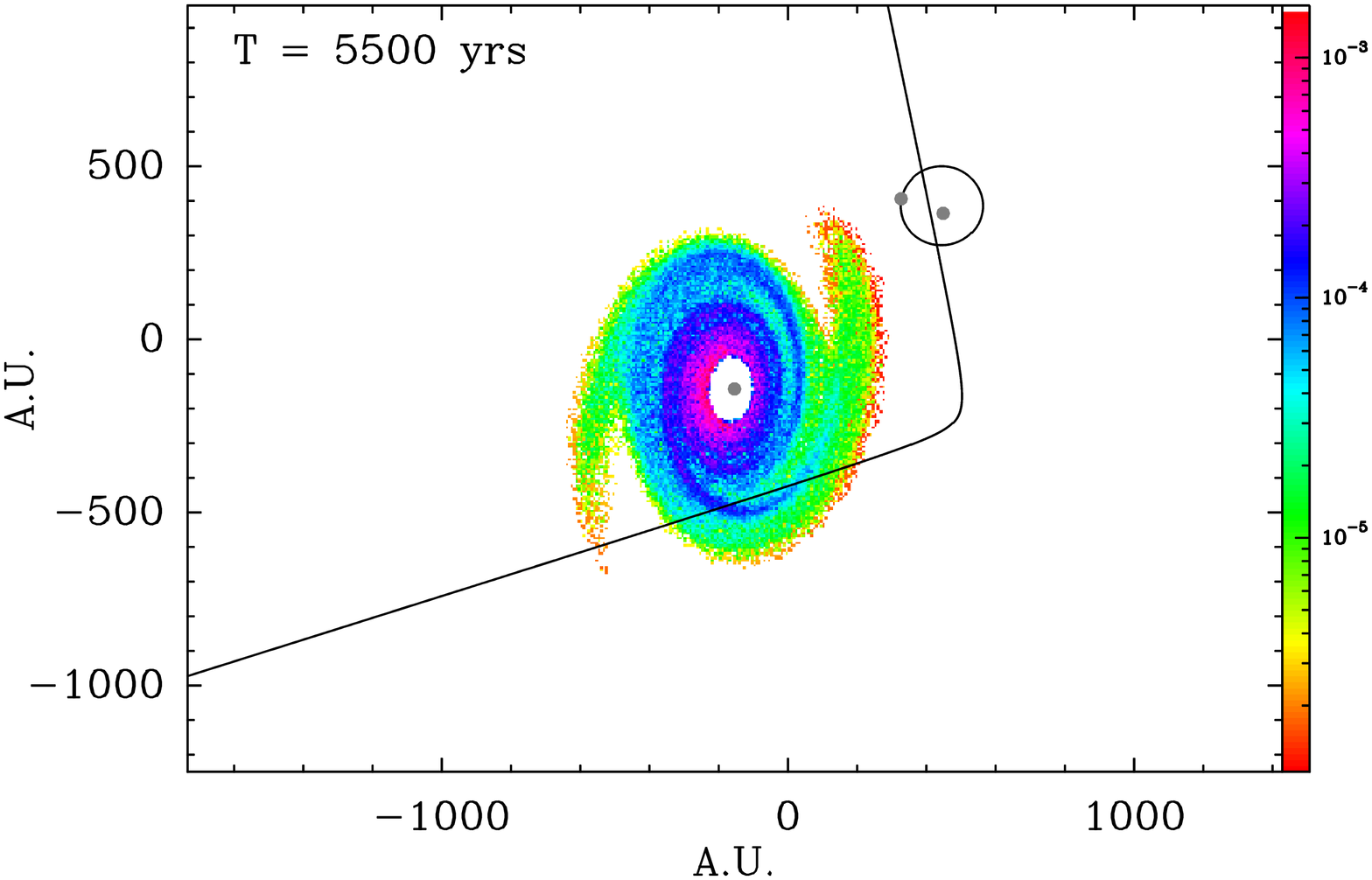}}\\ 
\caption{\label{contraintes} Examples of parameters for the
  flyby scenario inconsistent with the observations: the periapsis is
  too close to
  the disk, at $170$ AU, (top left panel) the crossing time at periapsis is too
  recent, $700$ years, (top right), the binary is still not yet
  at periapsis (bottom left) and the disk is too large (bottom
  right). These figures plot the intensity, in logarithm scale, of
  scattered light by the planetesimal spatial distribution obtained
  with our simulations. The scattering anisotropic factor is
  $0.2$. The time is relative to the crossing time at periapsis, $t_0$.}
\end{figure*}

Among the solutions consistent with the kinematic constraints (Sect
\ref{kinematics}), the
periapsis of the orbit, $d_0$, (i.e. the closest approach) can range between $10$
and several thousands of AU. If the flyby is however too close to the star, the disk can be
totally dissipated  or, at
least, can show a spiral structure which strongly differs from the
observations (Fig. \ref{contraintes} top left panel). We numerically
explored a broad range of $d_0$ values consistent with the kinematic
constraints of Sect. \ref{kinematics}. According to the simulations,
we find that 
the distance at periapsis, $d_0$ must be above about $600$ AU. An other
useful constraint is the time span between the periapsis passage
and the present time, $T=t-t_0$. If $T$  is too small compared to the keplerian
period at the outer disk edge, the spiral structure does not have the time to
develop (Fig. \ref{contraintes} top right panel). The simulations give an estimate
of at least $5\,000$ years before present for the periapsis passage.
All the well defined structures in the disk develop after the
periapsis passage, never before, which eliminates all the
configurations with $u<0$, because the binary is still
too far to generate any perturbations in the disk
(Fig. \ref{contraintes} bottom left panel). 

Few orbital configurations finally
verify the system given by the equations \ref{eq1} to \ref{vsky} plus
the additional constraints $d_0 > 600 \, AU$ and $T=t - t_0 > 5\,000 \, yrs$.
Valid solutions are only possible for $i$ between $85^\circ$ and $95^\circ$ and $u \approx
3$ and share the following properties: 
$$ 5\,000 \, \rm{yrs} < t - t_0 < 8\,000 \, \rm{yrs}$$
$$ 600 \, \rm{AU} < d_0 < 900 \, \rm{AU} $$
$$ e \approx 10 $$

  Thanks to Eq. \ref{irel}, it is also possible to determinate that:

$$35^\circ<i_{rel}<145^\circ$$

  Considering these solutions, the binary has at present time  already reached its
  velocity at infinity. Within this range of solutions,  the initial size of the
disk can also be  better constrain by our
simulations in the range between $400 - 450$ AU. If it is larger
(Fig.  \ref{contraintes} bottom right panel),
the flyby produces large spiral arms which are too extended while if
  the disk size is smaller, the disk does not fit the observed
  size. We do not address here the question of the origin of the disk
  size, nor the shape of the external disk edge. But collisional evolution of the disk \citep{2008AA...481..713T} or gravitationnal
  truncation  by the binary in a past bound state of the system 
  (see Sect. \ref{probability}) can both be used to explain them.

\begin{figure}
\centering
\includegraphics[width=0.48\textwidth]{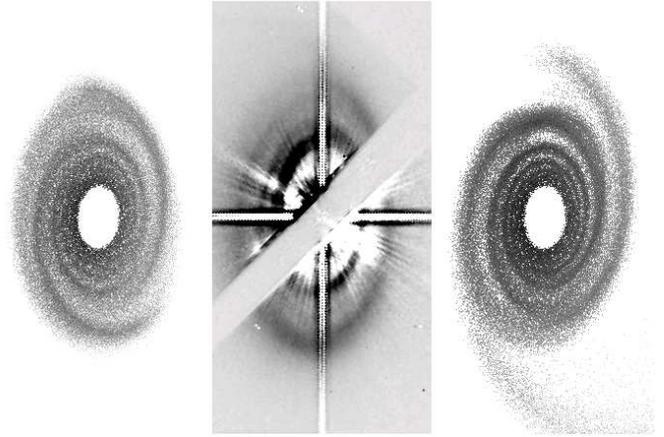}
\caption{\label{figure_flyby} Best solutions for the flyby scenario
  on a retrograde orbit (left panel) and on a prograde (right panel orbit, compared to
  the observations from \citet{2001A&A...372L..61M} (middle panel). A scattering anisotropic
  factor $g$ of $0.2$ is assumed to mimic real scattered light observations.}
\end{figure}

Overall, this scenario gives satisfactory results for the global shape of the
disk and reduces significantly the range of possible orbital
configurations for the companions. This is illustrated in
Fig. \ref{figure_flyby} which shows a disk of correct size with
a outer trailing spiral. The inner disk structure is nevertheless not
properly reproduced and the next step is thus to reproduce
more accurately the 
two  ring-like structure at distances of about $200$ and $325$ AU from
the star \citep{2001A&A...372L..61M,2003AJ....126..385C}. It
appears however that  none of the valid solutions for the flyby is able to well generate
the two rings. In the case of retrograde orbit ($85^\circ \leq i<90^\circ$), the flyby
generates a clear two arms spiral structure
(Fig. \ref{figure_flyby}, left panel). In the case  of prograde
orbit ($90^\circ<i \leq 95^\circ$), as studied for instance by
\citet{2003ApJ...592..986P},  the two arms are more wound and
more extended but it is still too
different from the observations (Fig. \ref{figure_flyby}, right panel). In the two
cases, the  flyby does
not perturb the inner part of the disk, below $200$ AU. It thus
cannot explain the inner ring, nor the observed depletion around $150$
AU. It means that we need to add some other hypothesis to the model.

\section{The flyby + planet scenario}
\label{planet}
We can conclude from Sect. \ref{constraints} that the sole presence of
the two star companions is not enough to  well
explain the observed disk properties summarised in Sect. \ref{observations},
especially the double-ring structure. The flyby can, at least, constrain the
initial disk size between  $400$ and $450$ AU but it is necessary to
use other hypothesis to improve the model and to try to better fit the
disk geometry. As the observations show a strong depletion in the dust
surface density below $150$ AU, a plausible hypothesis is the
presence of an unseen planet which generates this gap.

In this section, we consider the case where we  add a planet within
$150$ AU in our
numerical models.  The
planet is supposed to be coplanar with the disk and its main parameters
are its mass, its orbital semi-major axis and its orbital
eccentricity. The simulations take into account the mutual
gravitational interactions between the planet and the $3$ stars, as
well as the gravitational perturbation of the disk by the planet but
not the feedback of the disk on the planet: the planetesimals are
still considered as test particles. The system consisting of the
central star, the disk and the planet is first evolved during $5$
Myrs (the assumed age of the system), then the two star companions are
introduced to simulate the flyby with one of the valid set of orbital
parameters found in Sect.  \ref{constraints}.

\subsection{The inner ring}

\begin{figure}
\resizebox{\hsize}{!}{\includegraphics[angle=-90]{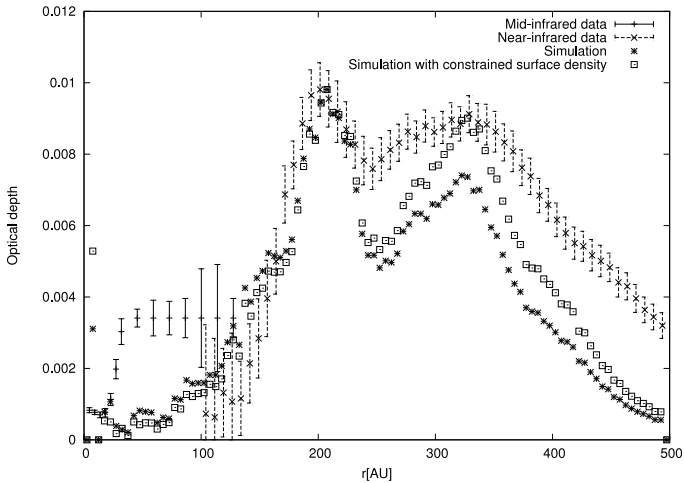}} 
\caption{\label{opticaldepth} Estimated profiles of dust optical depth
  (normal to the disk plane) by combining mid-infrared data
  \citep{2002ApJ...573..425M} and near-infrared data
  \citep{1999ApJ...525L..53W}, compared to the simulation output of
  the best scenario, a $2$ $M_J$ planet with an eccentricity of $0.2$.} 
\end{figure}

To constrain the planet parameters, we compare the simulations to the
 disk optical thickness derived from \citet{2002ApJ...573..425M}. The
 authors have combined their own observations in the mid-infrared and
 \citet{1999ApJ...525L..53W} observations in
the near-infrared  (Fig. \ref{opticaldepth}) to  estimate the radial
 profiles of the vertical dust optical
depth at $1.1$ $\mu$m. The observations in the mid-infrared cover the inner
 region of the disk, up to $130$ AU while the observations
 in the near-infrared cover the region between $130$ and $500$ AU. The
 radial profile of the optical depth at $1.1 \,\mu$m is deduced, for the
 mid-infrared data, from the optical depth at $12.5$ $\mu$m and
 assuming the opacity scales inversely with
 wavelength.

 The profiles show a ring which peaks around $200$ AU, a
depletion at $250$ AU and an even more depleted region below $150$
AU. We use in the following these data to constrain the mass and the orbital
elements of the planet, by trying to reproduce some key
characteristics of the azimuthal
averaged radial density profile of the dust:
\begin{itemize}
\item a peak between $190$ and $210$ AU
\item a ratio $\tau_{peak}/\tau_{250}$ of about $1.3$ between the peak
  surface density at $\sim 200$ AU and the depletion at $\sim 250$ AU
\item a ratio $\tau_{peak}/\tau_{150}$ of about $2.7$  between the
  peak at $\sim 200$ AU and the inner depleted region at
  $150$ AU. The uncertainty on this ratio is nevertheless rather
  large because of the large uncertainties in the optical depth
  between $100$ and $150$ AU.
\end{itemize}

The ``1 planet + flyby'' scenario is tested for $3$ different planet
masses: $0.2$, $2$ and
$8$ $M_{J}$. For each planet mass, $3$ apoapsis are considered ($130$,
$150$ and $180$ AU) with $4$ different values of the eccentricity: $0$,
$0.05$, $0.1$ and $0.2$. This gives a total of $36$ simulations and
their results are summarised in Tables \ref{peak}, \ref{rin} and
\ref{rout}. The peak position (Table \ref{peak}) indicates the
position of the largest over-density if several are
present. $\tau_{peak}/\tau_{150}$ is defined as the ratio between the
dust density  at the peak position
and the maximum density between $100$ and $150$ AU. $\tau_{peak}/\tau_{250}$ is
defined  as the ratio between the dust density at the peak position
and the density around $250$ AU. 

According to these simulations, a $0.2$ $M_J$ planet cannot
reproduce the observational data: it does not clear out enough the inner
part of the disk. For higher mass planets, the best solution is obtained for
an apoapsis of $130$ AU. For larger values of apoapsis, the peak position
is shifted toward too large distance from the star, or the planet can
even prevent the formation
of an over-density. For a $2$ $M_J$ planet, the best fits are obtained
for eccentricities around $0.2$, while for a $8$ $M_J$ planet, it must be
 below $0.1$. The best solution between these two planets is for a $2$ $M_J$ planet.

With an initial surface density distribution proportional to $r^{-1}$,
Fig. \ref{opticaldepth} shows that there is a lack of particles in the
outer part of the disk compared to the optical depth profile at
$\lambda=1.1 \,\mu$m. We therefore adjust the initial surface
density to obtain the same maximum at the outer peak location, near
$330$ AU. We find that the best fit is obtained for an initial surface
density  distribution
proportional to $r^{-0.5}$ and that the outer peak position in our
simulation is consistent with the observations. However, there is
still a lack of particles between the two rings and outside the outer
ring. This might reflect a limitation of our model as we only model the dynamics of
planetesimals. One can expect than smaller dust particles, sensitive
to radiation pressure and generated by collisions, will
migrate outside the rings and flatten the profiles, hence better reproducing the optical depth profile.

\begin{table*}
\caption{Summary of the results for the peak position (in AU) in the planetesimal
  density profile, for all the simulations, as a function of the planet
  mass, eccentricity and apoapsis. ``--'' indicates the simulations
  where no peak is observable or where the peak is below $170$ AU. $\dag$ indicates the simulations where
  multiple peaks appear. In this case, the largest peak is used to
  calculate the peak position, $\tau_{peak}/\tau_{150}$ and
  $\tau_{peak}/\tau_{250}$. The numbers boldface correspond to the
  best solutions.}
\label{peak}
\centering
\begin{tabular*}{\textwidth}{@{\extracolsep{\fill}}ll||ccccc|ccccc|cccc}
Planet mass&&\multicolumn{4}{c}{$0.2\;M_J$}&&\multicolumn{4}{c}{$2\;M_J$}&&\multicolumn{4}{c}{$8\;M_J$}\\
\cbrace{3-6}\cbrace{8-11}\cbrace{13-16}
Planet ecc.\rule[0pt]{0pt}{12pt}
&&$0$&$0.05$&$0.1$&$0.2$&&$0$&$0.05$&$0.1$&$0.2$&&$0$&$0.05$&$0.1$&$0.2$\\
\noalign{\smallskip}\hline\noalign{\smallskip}
apoapsis of $180$ AU&&$210$&$210$&$210$&--&&--&$230$&--&--&&--&--&--&--\\
apoapsis of $150$ AU&&$220\dag$&--&--&$230$&&--&$210\dag$&$230$&$240$&&--&--&--&--\\
apoapsis of $130$ AU&&--&--&--&$220$&&--&$220\dag$&-&$\mathbf{210}$&&$\mathbf{220}$&$\mathbf{220}$&$230$&$220$\\
apoapsis of $120$ AU&&--&--&--&--&&--&--&$200$&$\mathbf{190}$&&$\mathbf{220}$&$\mathbf{220}$&$230$&$230$\\
\end{tabular*}
\end{table*}

\begin{table*}
\caption{Same as Table \ref{peak}, but for the ratio $\tau_{peak}/\tau_{150}$
  between the density at peak position and the maximum density between
  $100$ and $150$ AU.}
\label{rin}
\centering
\begin{tabular*}{\textwidth}
{@{\extracolsep{\fill}}ll||ccccc|ccccc|cccc}
Planet mass&&\multicolumn{4}{c}{$0.2$ $M_J$}&&\multicolumn{4}{c}{$2$ $M_J$}&&\multicolumn{4}{c}{$8$ $M_J$} \\
\cbrace{3-6}\cbrace{8-11}\cbrace{13-16}
Planet ecc.\rule[0pt]{0pt}{12pt}
&&$0$&$0.05$&$0.1$&$0.2$&&$0$&$0.05$&$0.1$&$0.2$&&$0$&$0.05$&$0.1$&$0.2$\\
\noalign{\smallskip}\hline\noalign{\smallskip}
apoapsis of $180$ AU&&$0.5$&$0.5$&$0.5$&--&&--&--&$0.4$&--&&--&--&--&--\\
apoapsis of $150$ AU&&$0.5\dag$&--&--&$0.8$&&--&$0.6\dag$&$1.6$&$3.0$&&--&--&--&--\\
apoapsis of $130$ AU&&--&--&--&$1.1$&&--&$2.4\dag$&-&$\mathbf{2.3}$&&$\mathbf{3.0}$&$\mathbf{5.0}$&$6.0$&$8.0$\\
apoapsis of $120$ AU&&--&--&--&--&&--&--&$1.3$&$\mathbf{2.1}$&&$\mathbf{2.7}$&$\mathbf{4.7}$&$4.2$&$4.5$\\
\end{tabular*}
\end{table*}

\begin{table*}
\caption{Same as in Table \ref{peak}, but for the ratio $\tau_{peak}/\tau_{250}$
  between the density at peak position and the density in the gap
  around $250$ AU.}
\label{rout}
\centering
\begin{tabular*}{\textwidth}{@{\extracolsep{\fill}}ll||ccccc|ccccc|cccc}
Planet mass&&\multicolumn{4}{c}{$0.2$ $M_J$}&&\multicolumn{4}{c}{$2$ $M_J$}&&\multicolumn{4}{c}{$8$ $M_J$} \\
\cbrace{3-6}\cbrace{8-11}\cbrace{13-16}
Planet ecc.\rule[0pt]{0pt}{12pt}
&&$0$&$0.05$&$0.1$&$0.2$&&$0$&$0.05$&$0.1$&$0.2$&&$0$&$0.05$&$0.1$&$0.2$\\
\noalign{\smallskip}\hline\noalign{\smallskip}
apoapsis of $180$ AU&&$1.5$&$1.7$&$1.3$&--&&--&--&$1.1$&--&&--&--&--&--\\
apoapsis of $150$ AU&&$1.7\dag$&--&--&$1.8$&&--&$1.4\dag$&$1.5$&$2.0$&&--&--&--&--\\
apoapsis of $130$ AU&&--&--&--&$2.0$&&--&$1.7\dag$&--&$\mathbf{2.0}$&&$\mathbf{1.9}$&$\mathbf{1.9}$&$1.7$&$1.3$\\
apoapsis of $120$ AU&&--&--&--&--&&--&--&$2.2$&$\mathbf{2.1}$&&$\mathbf{1.8}$&$\mathbf{1.7}$&$1.8$&$1.4$\\
\end{tabular*}
\end{table*}

\subsection{The global structure}

\begin{figure*}
\centering
\includegraphics[width=\textwidth]{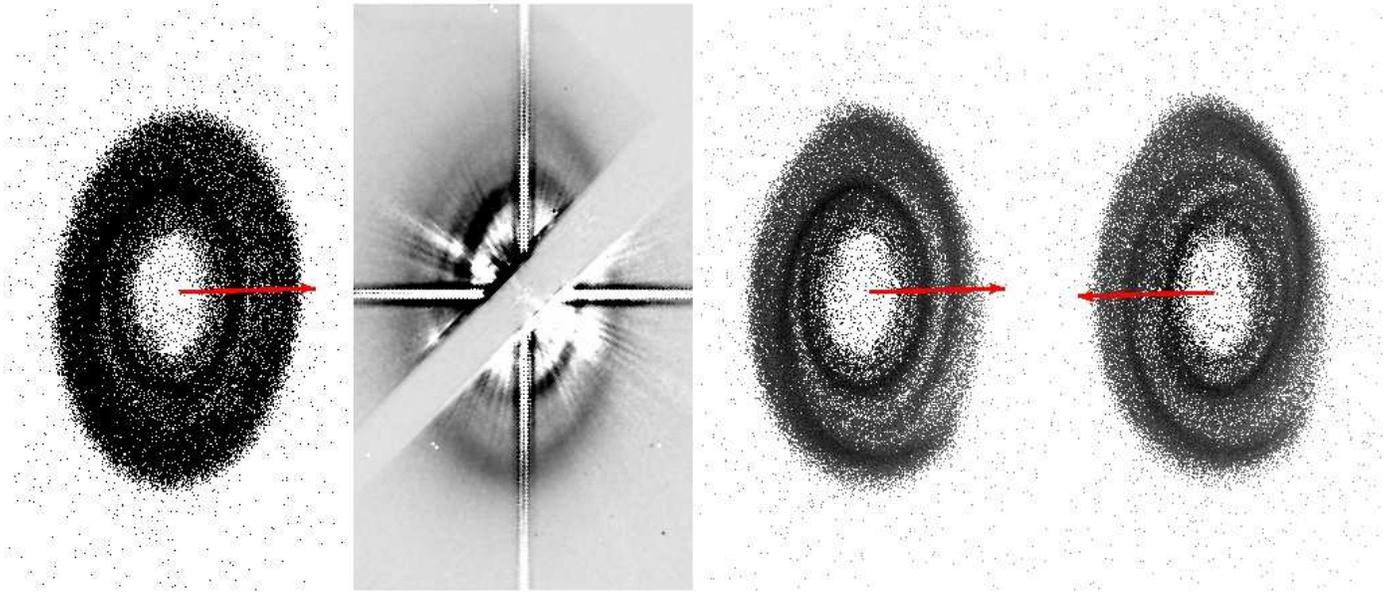} 
\caption{\label{figure_flyby_planet} Same as in
  Fig. \ref{figure_flyby}, but for the best solution for the flyby +
  planet scenario in the case of a retrograde orbit of the binary
  (third panel). The red arrow indicates the direction of the planet
  periapsis. The
  planet mass is equal to $2$ $M_J$, it apoapsis is $130$ AU and its
  eccentricity $0.2$. The first panel show the disk before the flyby
  but after $5$ Myrs of gravitational perturbation by the planet. The
  fourth panel shows an alternative configuration where the initial
  direction of the planet periapsis is turned by $180^\circ$.}
\end{figure*}

The ratio between the
two peaks in the optical depth profile depends of the observing wavelength, as
the optical depth at the outer peak position is larger than the
optical depth at the inner peak position at $0.5$ $\mu$m
\citep{2005ApJ...627..986A}, while it is the opposite of
Fig. \ref{opticaldepth} ($\lambda=1.1 \,\mu$m). Our constraint
on the initial surface density distribution should therefore be taken
with care, but our model is nevertheless very successful to
predict the position of the outer peak in the optical depth
profile. The planet parameters are strongly constrained by the inner
ring, but our simulations show that the planet also perturb the outer
parts of the disk as shown in Figs. \ref{opticaldepth} and
\ref{figure_flyby_planet}. 

In less than
$5$ Myrs, the planet has time enough to develop a strong,
one-armed, anticlockwise wound spiral, due to the differential precession rate of the
planetesimal orbits as explained by \citet{2004AA...414.1153A}. The
flyby does  thus not occur with an initially
smooth disk and this has strong consequences on the final shape of
the structures produced by the flyby. For instance, in the case where
the binary is on a retrograde orbit, we previously obtained a regular
two armed spiral. But now, the planet not only produces the inner
ring, but can also make disappear, or at least perturb,  a significant
fraction of one of the two
spiral arms. As shown in Fig. \ref{figure_flyby_planet}, the
  disk morphology between $200$ and $300$ AU slightly depends of the
  direction of the planet periapsis with respect to the direction of the binary
  periapsis, but the differences remain too small to constrain the
  orientation of the planet orbit. The results are thus not perfect
because the simulations do not still reproduce exactly the outer ring
but they clearly show that we need a complete description of the
system to properly account for it. It is not satisfactory to consider solely the external
companions or the inner planet. Combining both, we clearly achieve a
much better fit to the observations.

\section{The flyby + 2 planets scenario}
\label{2planets}

\begin{figure}
\centering
\includegraphics[width=0.48\textwidth]{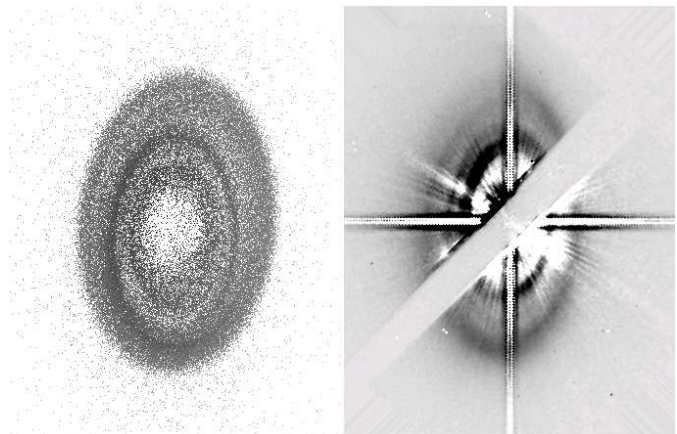} 
\caption{\label{figure_2planets} Same as in
  Fig. \ref{figure_flyby} , but for the flyby +
  2 planets scenario. The
  inner planet mass is equal to $2$ $M_J$, it apoapsis is $130$ AU and its
  eccentricity $0.2$. The outer planet mass is $0.2$ $M_J$, it apoapsis is $250$ AU and its
  eccentricity $0.05$.}
\end{figure}

As the previous scenario is still not fully satisfactory because it
still does not generate two rings clearly separated by a large gap, we
have  also investigated an
alternative with two planets in the disk. \citet{2005DPS....37.2810W}
has already considered the case of a single planet at about $250$ AU to
explain the outer ring but without considering the perturbation due to the
companion stars.  We
therefore first investigate  the consequence of both the companions and the
planet on the disk geometry. Using the solutions found in Sect.
\ref{constraints}, it appears that adding the gravitational influence
of the two companion stars does not improve the structure generated by
the external planet but, on contrary, disrupts them because the
planetesimals orbits are too destabilised  by the perturbations of a
planet and that of the two companions.  For this scenario,
the constraints used in Sect. \ref{geometry} are no longer valid and
the correct orbital elements for the binary orbit are those which prevent interactions
between the disk and the two M stars: they do not still reach the
periapsis location ($u < 0$) or the periapsis itself is located at more than
$1500$ AU from the primary star.

Such a scenario cannot, however, reproduce the inner
depletion below $150$ AU; another planet is thus necessary like in the
previous section. According to \citet{2005DPS....37.2810W}, two
scenarios can be considered: a $0.2$ $M_J$ planet
evolving over $5$
Myrs on an orbit with a semi-major axis of $250$ AU and an eccentricity
of $0.05$, and a $2$ $M_J$ planet perturbing the
disk over $0.5$ Myrs and on the
same orbit. We add thus another planet in the
internal part of the disk, following the best solutions
found in the previous section, namely a $2$ $M_J$ planet on an orbit with a
semi-major axis of $110$ AU and an eccentricity of $0.2$ or a $8$
$M_J$ planet  on an orbit with a semi-major axis of $115$ AU and an
eccentricity of $0.05$ (Fig. \ref{figure_2planets}).

\begin{figure}
\resizebox{\hsize}{!}{\includegraphics[angle=-90]{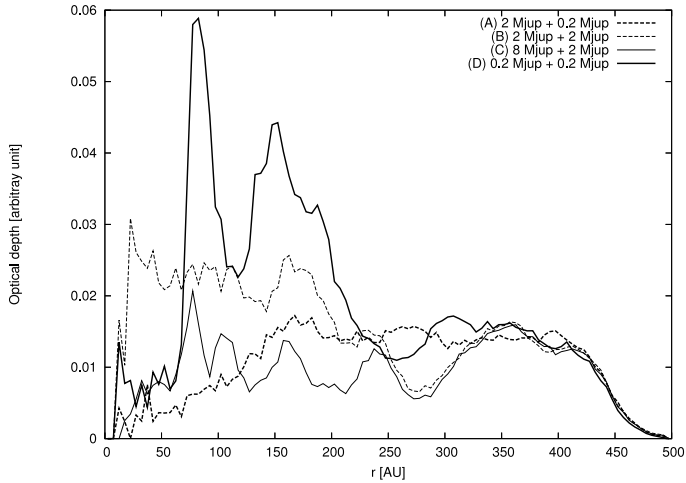}} 
\caption{\label{depth} Optical depth obtained by simulations with $4$
  different scenarios for the mass of the $2$ planets, to compare to
  Fig.\ref{opticaldepth}.} 
\end{figure}

Like in Sect. \ref{planet}, the estimated radial profiles of the dust optical
depth obtained with these simulations are compared to the observed
optical depth at $1.1$ $\mu$m (Fig. \ref{opticaldepth}). Two major  informations are therefore given by this
method. First, and independently of the mass of the inner planet, a $2$ $M_J$ at $250$ AU generates a too deep gap in
the disk compared to the observations (scenario B and C of
Fig. \ref{depth}). Contrary to scenario ``flyby + 1 planet'', one
cannot expect that the optical depth profile would be flattened by
smaller dust particles sensitive to radiation pressure because the
particles will be quickly 
removed from the gap, due to close
encounter with this planet orbiting at $250$ AU. \citet{2005DPS....37.2810W} constrains the planet parameters
considering the location and the winding of the spiral but there is a
degeneracy between the planet mass and the time the planet is
orbiting around the star. If one adds the constrain given by the
observations on the depth of the gap around $250$ AU, only the
scenarios with smaller planet masses ($0.2$ $M_J$) orbiting for a
long time ($5$ Myrs) are satisfactory (scenario A and D of
Fig. \ref{depth})). Second, the simulations show
that it is difficult to obtain an inner ring massive enough between
the two planets, around $200$ AU. Due to the perturbations of these two objects, a lot of
planetesimals are ejected from the disk. We have therefore tried other
simulations with a smaller mass ($0.2$ $M_J$, scenario D) for the internal
planet in order to less perturb the disk. However, the results are
still not satisfactory because the peak position is not at the correct
location ($150$ AU instead of $200$ AU) and because, with a low mass
planet, there are too much planetesimals below $100$ AU.

A planet located at $250$ AU from the primary star is a interesting
alternative to the companions in order to explain the outer ring
without the gravitational perturbation of the $2$ stellar companions, but
our work shows that combining it with an other planet around $120$ AU
to explain the inner ring and the dust depletion within $150$ AU does
not provide satisfactory results. Even if
it is still not perfect, the scenario ``flyby + 1 planet'' is  the
one which gives the best results among the three studied scenarios in this paper.

\section{Discussion}
\label{discussion}

\subsection{Probability of the encounter}
\label{probability}
The main drawback to the flyby scenario is that, according to our
models, the binary passed to the periapsis less than $8000$ years
ago. The probability of observing such an event is low, making the
flyby scenario statistically unlikely. On the other hand, the similar
physical properties between the central star and the low mass
companions suggest a common origin. The $3$ stars have
been independently estimated to be around $5$ Myrs
\citep{2000ApJ...544..937W,2004A&A...419..301M} and the differential
radial velocity between the HD 141569A and the binary is only a fraction of
the total speed of the primary star. But the observed
radial velocities clearly show that the binary is on an escape
orbit and the open spiral structure observed in the disk is
also in favour of a flyby scenario, as trailing structures are created
by tidal effects (Augereau et al., in prep.). 

One can however reconciliate these positions with an hybrid
scenario. HD 141569 could have been, in the past, a triple bound system until
an other star of its vicinity approached and destabilised the
system, putting the binary on its current orbit. In this case, we
would be observing the last
periapsis passage of the binary and the consequences of a $4$ star
encounter. This scenario could explain the observed velocities, the
common physical properties and the disk size, and it is more plausible that a sole flyby
of the binary. The gravitationally bound scenario studied by
\citet{2004AA...414.1153A} and \citet{2005AJ....129.2481Q} can possibly be used
to constraint the orbit of the binary before the encounter: typically
an orbital period of several hundreds of thousands years and a
apoapsis of almost $10000$ AU. It appears more probable that a fourth
star destabilises such an orbit that the binary randomly cross the path
of the primary star at less than $1000$ AU. The study done by \citet{2001MNRAS.325..449S}
show that, in the Orion Nebula Cluster, almost all the stars undergo
at least one encounter between $1000$ and $10000$ AU. On the other
hand, \citet{2008arXiv0809.3289A} have found
that HD 141569 formed in relative isolation, tens of parsecs away from
the recent sites of star formation in the Ophiucus-Scorpius-Centaurus
region. The likehood of such a scenario has thus to be better estimated.
Its dynamical modelling has also to be investigated by this is beyond
the scope of this paper. Theoretical works
are however already abundant in the literature (see, for example, the series
discussing many aspects of the three-body scattering from
\citet{1983ApJ...268..319H} to \citet{1996ApJ...467..359H}).

\subsection{Comparison with previous works}

Other theoretical studies of the \hd system can be
divided in $3$ groups with respect to the reproduced disk structures:
the outer ring, the gap between the two rings or the extended spiral arms.

\citet{2004AA...414.1153A} and \citet{2005AJ....129.2481Q} have both studied the
outer ring of the disk, but using different tools: a pure N-body code
for \citet{2004AA...414.1153A} and a 2D hydrodynamics code for
\citet{2005AJ....129.2481Q} assuming in the latter case that the gas
and the dust are fully mixed. Their main issue compared to our work is that
they have considered bound orbits for the binary which is incompatible
with the
new measurement of radial velocities for the primary star. Interestingly,
\citet{2004AA...414.1153A} well reproduce the outer ring but not the
open spiral arms structure while it is the contrary for our work. One
can therefore ask if the scenario with $4$ stars can reconciliate
these two approaches. This has to be investigated in a future paper.

\citet{2005DPS....37.2810W} and \citet{2001ApJ...557..990T} have considered the
opening of the gap between the two rings with very different
mechanisms: a planet embedded in the disk \citep{2005DPS....37.2810W}
or the sole interaction between the dust and the gas
\citep{2001ApJ...557..990T}. In the two cases, the authors do not consider
the influence  of the two companion stars which can be justified by
our study of the kinematic constraints. It is indeed  possible
to obtain a binary orbit reproducing the observed radial velocities
but which does not let the two M stars generate any perturbation to
the disk because they pass too far from the disk or because they have
not yet passed at the periapsis. The planet used by
\citet{2005DPS....37.2810W} is efficient to open the gap and generate
the outer ring but, as shown in Sect. \ref{2planets}, it is difficult
to improve this scenario in order to also explain the inner
ring and the inner disk depletion. \citet{2001ApJ...557..990T}, with
an alternative scenario of dust
migration due to the gas friction, generate a two ring
structures but the hypothesis of axysmetricity does not leave a lot of
possibilities to extend this work and explain the tightly-wound spiral
structure of the outer ring nor the spiral arm in the extended diffuse
emission.

\citet{2005ApJ...627..986A} consider several physical processes for the
modelling of \hd: gravitation, interaction between the gas and
the dust, dust generation by collisions. It is therefore, up to now, the most complete
theoretical study about \hd.  They considered the case of a flyby
in order to obtain large spiral arms similar to those observed in the
extended diffuse emission. They also added a  $5$ $M_J$ planet, on
an orbit with a semi-major axis of $100$ AU with an eccentricity of
$0.6$, to explain the depleted zone below $150$ AU. Altought the
authors obtain good results for the  the extended spiral arms and
the inner hole, they did not succeed to reproduce the two rings nor the gap
in between. One possible explanation is that, in our work, we obtain
satisfactory results only if we let the disk evolves with the
planet during several million years before the flyby. To go
further in the description of the \hd  system, it will
be therefore interesting to use the model developed by these authors
but with the initial conditions on the binary orbit and on the planet
parameters established by our work.

\section{Summary and conclusion}
\label{conclusion}

The binary orbit is finally found to be almost fixed by the
   observational constraint on a edge-on plane with respect to the
   observer. If the binary has had an influence on the disk
   structure, it should have a  passing time at the periapsis between $5000$
   and $8000$ years ago and a distance at periapsis between $600$ and
   $900$ AU. The scenarios with retrograde orbits better reproduce
   the observations but prograde orbits cannot be totally excluded. In order to reproduce the observed structures
   in the debris disk, it also appears that the best scenario is a
   flyby with $1$ planet embedded in the disk. For a $2$ $M_J$
   planet, its orbital eccentricity must be around $0.2$ while for a $8$
   $M_J$ planet, it must be below $0.1$. In the two cases, its
   apoapsis is about $130$ AU.

This scenario is able to both reproduce the observed radial
velocities of the stars and the overall structure of the disk with the
modelling of a flyby by the binary and a embedded planet in the
disk. However, this scenario is not fully satisfactory because it does
not reproduce all the structures, and in particular the entire gap
between the two annulus. Future study can therefore extend
this scenario, for example with the modelling of the gas and/or
collisions, as we have not taken into account these effects on the dust dynamics.

\begin{acknowledgements}
We are grateful to Ken Marsh for the optical depth observational data. We thank Xavier Delfosse for the discussion about the flyby encounter
probability. We also
thank John Papaloizou and the anonymous referee for helpful comments on this paper. Most of the
computations presented  in this paper were performed at the Service
Commun de Calcul  Intensif de l'Observatoire de Grenoble (SCCI).
\end{acknowledgements}

\bibliographystyle{aa}
\bibliography{biblio}

\end{document}